\documentclass[aps,prd,groupedaddress,preprint,eqsecnum,nofootinbib]{revtex4}
\usepackage{graphicx,epsf,amssymb,amsbsy,amsfonts,amssymb,amsmath}
\begin{document}
\hfuzz 12pt

\title{RG Flow and Thermodynamics of Causal Horizons in AdS}
\author{Shamik Banerjee}
\affiliation{Kavli Institute for the Physics and Mathematics of the Universe, The University of Tokyo,\\
5-1-5 Kashiwa-no-Ha, Kashiwa City, Chiba 277-8568, Japan}

\email{banerjeeshamik.phy@gmail.com}

\begin{abstract}
Causal horizons in pure Poincare $AdS$ are Killing horizons generated by dilatation vector. Renormalization group (RG) flow breaks the dilatation symmetry and makes the horizons dynamical. We propose that the boundary RG flow is dual to the thermodynamics of the causal horizon. As a check of our proposal we show that the gravity dual of the boundary $c$-theorem is the second law of thermodynamics obeyed by causal horizons. The holographic $c$-function is the Bekenstein-Hawking entropy (density) of the dynamical causal horizon. We explicitly construct the $c$-function in a generic class of RG-flow geometries and show that it interpolates monotonically between the UV and IR central charges as a result of the second law. 

\end{abstract}

\preprint{IPMU15-0140}
\maketitle

\section{Introduction}
Renormalization group flow is an important aspect of any field theory. In holography the additional bulk direction(s) is said to emerge from the renormalization group flow of the dual field theory. This connection has beed made manifest in holographic renormalization group flow studied, e.g, in \cite{Akhmedov:1998vf, Freedman:1999gp, Girardello:1998pd, deBoer:1999xf, Myers:2010xs, Myers:2010tj}. One of the cornerstones of RG-flow is its monotonicity property exemplified by $c$-theorem \cite{Zamolodchikov:1986gt, Cardy:1988cwa, Osborn:1989td, Komargodski:2011vj, Komargodski:2011xv, Casini:2004bw, Casini:2012ei}. Holographic $c$-functions have been constructed before as radial evolution of a locally constructed function of the metric. In this paper we would like to address this problem from a different perspective. We will see that (thermo)dynamics of causal horizons play a crucial role in this.

\section{Dilatation and causal horizons in ADS}

\begin{figure}[htbp]
\begin{center}
  \includegraphics[width=15cm]{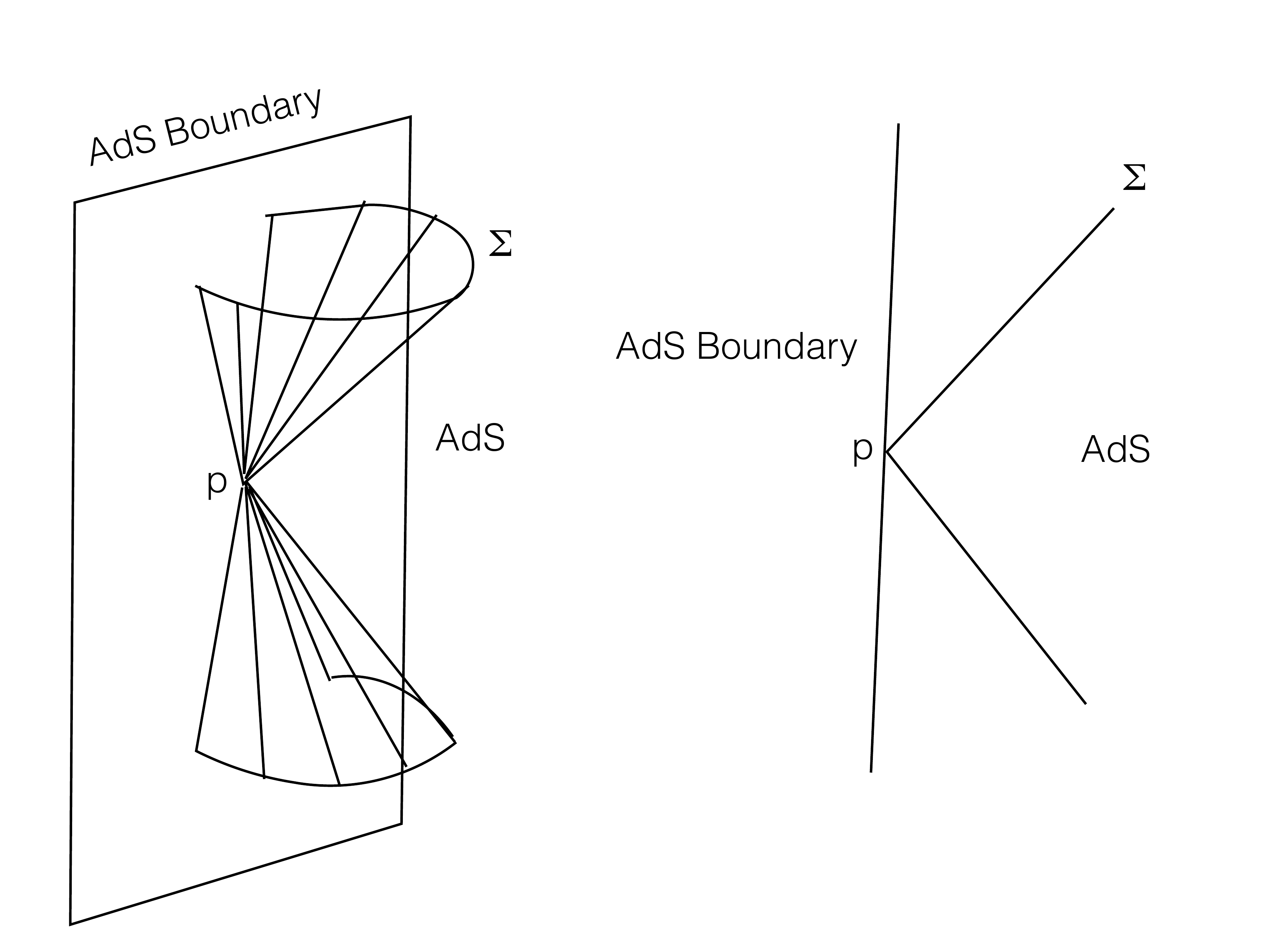}
\end{center}
\caption{Causal horizon $\Sigma$  of the boundary point p in pure AdS}
\end{figure}


We want to study AdS-CFT in the Poincare patch so that our field theory lives in Minkowski space-time. In the Poincare patch the metric of $AdS_{d+1}$ can be written as, 

\begin{equation}
ds^2 = \frac{L^2}{z^2} \big(\eta_{\mu\nu}dx^\mu dx^\nu + dz^2 \big)
\end{equation}
where $L$ is the AdS radius and $\eta_{\mu\nu}$ is the $d$-dimensional Minkowski metric with mostly $+$ entries. This metric has a scaling isometry, $(x^{\mu},z \rightarrow \lambda x^{\mu}, \lambda z)$, generated by the vector field,

\begin{equation}
D = x^\mu \frac{\partial}{\partial x^\mu} + z\frac{\partial}{\partial z}
\end{equation}

The isometry generated by the vector $D$ has the interpretation of the scaling symmetry of the conformal field theory living on the boundary of $AdS$. 

The norm of $D$ is given by, 

\begin{equation}
D^2 = \frac{L^2}{z^2} \big(\eta_{\mu\nu}x^\mu x^\nu +z^2 \big)
\end{equation}

This is a causal vector in the region, $D^2\le 0$. In particular $D$ is null on the hypersurface $\Sigma$ given by, 

\begin{equation}
D^2 =0 = \eta_{\mu\nu}x^\mu x^\nu +z^2
\end{equation}

$\Sigma$ is a null-hypersurface which is left invariant by the flow generated by $D$. Since $D$ is Killing and null on $\Sigma$, $\Sigma$ is a Killing horizon. It can be easily checked that it is non-degenerate.

$\Sigma$ also has another interpretation. If we choose the boundary point $p$ given by the coordinates $x^{\mu}=z=0$, then $\Sigma$ is the boundary of the causal (future)past of $p$ in the $AdS$ bulk. This is also the boundary of the causal (future)past of a time-like curve in $AdS$ with infinite proper length in the (past) future and ends at the point $p$. So this null hypersurface $\Sigma$ is the causal horizon of the bulk observer who reaches the point $p$ on the boundary. 

As we have shown the causal horizon is a Killing horizon in pure $AdS$ generated by the dilatation $D$.  Now suppose we deform the boundary CFT by adding a relevant operator. This breaks the scale invariance and causes renormalization group flow.  In the deep IR the field theory reaches a fixed point. The IR fixed point could be some non-trivial conformal field theory or could be some theory with mass gap. In any case, the dual gravity description will not have any scaling symmetry globally. But since we are adding only relevant operator the bulk still remains asymptotically $AdS$ and the scaling symmetry is recovered asymptotically. 

Deforming the UV-CFT by adding relevant operator is dual to the addition of massive scalar fields with negative mass squared in the bulk. The backreacted geometry is only asymptotically $AdS$. Now the causal horizon of the observer reaching the boundary point $p$ shifts due to the backreaction of the matter. In particular this is no longer a Killing horizon - the dilatation is broken. The horizon will be Killing near the point $p$ because scaling is recovered asymptotically. This is similar to what happens if we perturb a black hole such that the perturbation vanishes in the far future. The black hole settles down to a stationary state as the effect of the perturbation dies away. In particular there are laws of thermodynamics which govern the future evolution of the perturbed horizon. In our case the role of the black hole horizon is played by the causal horizon of the bulk observer reaching the boundary point $p$. The causal horizon fluctuates because there is a non-zero flux of energy-momentum tensor of the matter through the horizon. The matter is the scalar field dual to the operator which we have used to deform the UV-CFT. The causal horizon becomes stationary near the boundary of $AdS$ because the effect of a relevant perturbation vanishes in the UV. This is a (past)future boundary condition from the point of view of the causal horizon and is dual to the restriction that we do not add irrelevant operator to the field theory because it destroys the UV fixed point. The time evolution of the causal horizon also satisfies laws of thermodynamics in particular the second law which states that the area of (past)future causal horizon does not (increase)decrease \cite{Gibbons:1977mu, Jacobson:2003wv}. This is a natural monotonicity property associated with a dynamical causal horizon and is reminiscent of $c$-theorem. 

So the basic idea is to replace the Poincare time by dilatation time in the bulk. Although there is no black hole horizon, causal horizons exist and they are Killing with respect to the dilatation-time. The structure of empty AdS resembles that of a stationary black hole space-time if we use dilatation time. The dilatation vector is not globally time-llike. It is time-like outside the causal horizon, space-like behind the causal horizon and null on the horizon. In this set up conformal field theories represent the equilibrium configurations and non-conformal field theories represent the non-equilibrium configurations. RG flow is the dilatation-time evolution in the bulk which takes the non-equilibrium system to the final equilibrium configuration. So the gravity dual of the boundary RG flow is a thermalization process where now the role of the more conventional black hole horizon is being played by causal horizons. $c$-function plays the role of thermal entropy in this picture.

More concretely we would like to propose that the dynamics of the causal horizon due to the breaking of the scaling symmetry is dual to the renormalization group flow in the field theory. As a non-trivial check of this we will show that the $c$-theorem in the field theory is dual to the second law of thermodynamics of the causal horizon, with the Bekenstein-Hawking entropy(density) of the dynamical causal horizon playing the role of the $c$-function.

\section{c-theorem and second law of thermodynamics for causal horizon}

\begin{figure}[htbp]
\begin{center}
  \includegraphics[width=15cm]{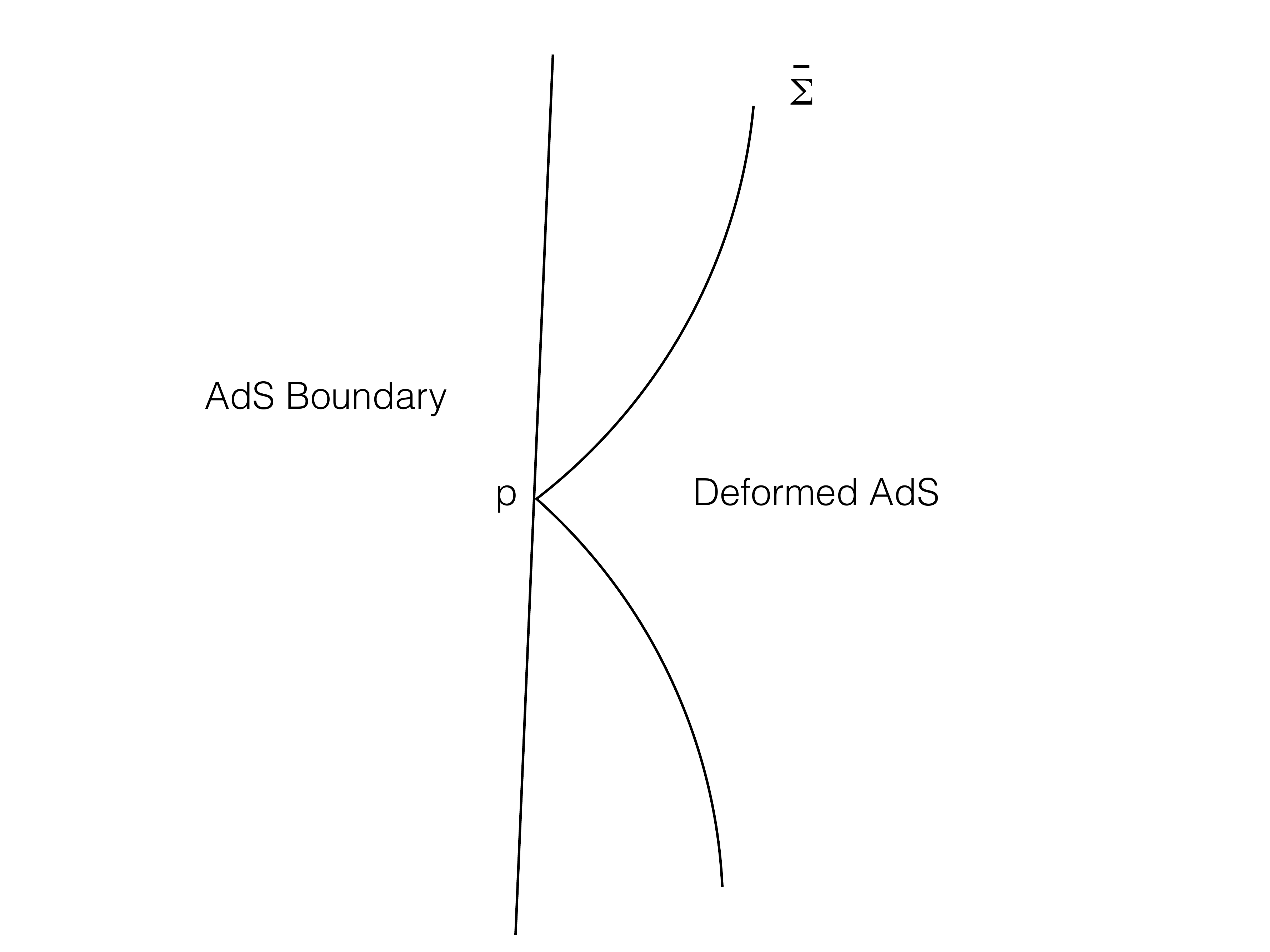}
\end{center}
\caption{Deformed causal horizon $\bar\Sigma$ of the boundary point p in deformed AdS}
\end{figure}

For the sake of simplicity, let us start from $AdS_3$ space. In pure $AdS_3$ the causal horizon of the boundary point $p$ with coordinates $x^\mu=z=0$  is given by, 

\begin{equation}
-t^2 + x^2 + z^2 =0 
\end{equation}

We will work with the past horizon, denoted by $\Sigma$, i.e, the boundary of the causal future of the point $p$. This has the property that its area cannot increase in the future direction. This is closer to the conventional statement of the $c$-theorem which states that the $c$-function is monotonically decreasing as we move from the UV to the IR. 

Now suppose we deform the boundary conformal field theory by adding a relevant operator which corresponds to adding a massive scalar field in the bulk. The back reaction of the matter will change the pure AdS geometry. Let us take the form of the back reacted geometry to be, \footnote{See for example \cite{Liu:2012eea}.}

\begin{equation}
ds^2 = \frac{L^2}{z^2} \Big(-dt^2 + dx^2 + \frac {dz^2}{f(z)}\Big)
\end{equation}

where $f(z)$ has the form determined by solving the coupled Einstein-Scalar equations. The scalar field is assumed to satisfy null energy condition and this leads to the constraint that, 

\begin{equation}
f'(z) \ge 0
\end{equation}

We also have, $f(z)\rightarrow 1$ as $z\rightarrow 0$. The behavior of $f(z)$ as $z\rightarrow \infty$ depends on whether the IR is a non-trivial CFT or a gapped theory. If there is a non-trivial interacting CFT in the IR then, the behavior of $f(z)$ for large $z$ is given by,

\begin{equation}
f(z)\rightarrow \frac{L^2}{L_{IR}^2}
\end{equation}

where $L_{IR}$ is the size of the locally $AdS$ region in the deep interior. If instead the IR theory is gapped then $f(z)$ blows up as $z\rightarrow \infty$. 

Let us now define the new variable $\bar z$ by, 

\begin{equation}\label{bar}
\bar z = \int_{0}^{z} \frac{dz'}{\sqrt {f(z')}}
\end{equation}

With this new variable the equation of the perturbed metric becomes, 

\begin{equation}
ds^2 =  \frac{L^2}{z^2} \big(-dt^2 + dx^2 + d\bar z^2 \big)
\end{equation}
where we treat $z$ as a function of $\bar z$. The equation of the causal horizon, $\bar \Sigma$, in the perturbed geometry becomes, 

\begin{equation}\label{d-null}
-t^2 + x^2 + \bar z^2 =0 
\end{equation}

Now we define a vector $\bar D$ as,

\begin{equation}
\bar D = t \frac {\partial}{\partial t} + x \frac {\partial}{\partial x} + \bar z \frac {\partial}{\partial \bar z}
\end{equation}

$\bar\Sigma$ is invariant under the flow generated by $\bar D$ and $\bar D$ is null on $\bar\Sigma$. So $\bar D$ is the null generator of $\bar\Sigma$. Asymptotically $\bar D$ becomes Killing and generates the scaling symmetry of the UV-CFT. 

Let us take the null generators to be the integral curves of $\bar D$ parametrized by $\lambda$, 

\begin{equation}\label{intcurve}
\frac{dt}{d\lambda} = t , \\ \frac{dx}{d\lambda} = x, \\ \frac{d\bar z}{d\lambda} = \bar z
\end{equation} 

$\lambda$ is not necessarily an affine parameter. The components $(t(\lambda),x(\lambda),\bar z(\lambda))$ satisfy $(\ref{d-null})$,

\begin{equation}
-t(\lambda)^2 + x(\lambda)^2 + \bar z(\lambda)^2 =0
\end{equation}

Let us now parametrize $\bar\Sigma$ by writing,

\begin{equation}
t = \bar z \cosh{\eta}, \\ x= \bar z \sinh{\eta}, \\ \bar z = \bar z
\end{equation}

So $(\bar z, \eta)$ are intrinsic coordinates on $\bar\Sigma$. Now along the null generators we can write, 

\begin{equation}\label{ng}
t(\lambda) = \bar z(\lambda) \cosh{\eta(\lambda)}, \\ x(\lambda)= \bar z(\lambda) \sinh{\eta(\lambda)}
\end{equation}

$(\bar z(\lambda), \eta({\lambda}))$ is the equation of the null generators in the intrinsic coordinates. It is easy to see using ($\ref{intcurve}$) and ($\ref{ng}$) that $\eta(\lambda)$ is independent of $\lambda$. In other words $\eta$ is a comoving coordinate and parametrize a spatial slice of $\bar\Sigma$.

The induced metric on $\bar\Sigma$ is given by, 

\begin{equation}
ds_{ind}^2 = L^2 \frac{\bar z^2}{z^2} d\eta^2
\end{equation}

Now, 

\begin{equation}
\frac{d}{d\lambda} \Big(L\frac{\bar z}{z}\Big)= L \frac{\bar z}{z} \int_0^{1} \Bigg(1 - \frac{\sqrt {f(z(\lambda))}}{\sqrt {f(\alpha z(\lambda))}}\Bigg) d\alpha
\le 0
\end{equation}

because $f(z)$ is a monotonically increasing function of $z$. We also take note of the fact that $\eta$ being a comoving coordinate, $d\eta$ does not depend on $\lambda$.

Now the Bekenstein-Hawking entropy density is given by, 

\begin{equation}
\frac{dS_{BH}}{d\eta} = \frac{1}{4G_N} \Big(L\frac{\bar z}{z}\Big)
\end{equation}

and satisfies,

\begin{equation}
\frac{d}{d\lambda} \frac{dS_{BH}}{d\eta} \le 0
\end{equation}

This is the statement of $c$-theorem. To see this more clearly let us note that increasing $\lambda$ corresponds to moving away from the boundary point $p$. In particular let us use equation (\ref{bar}) to write,

\begin{equation}\label{bar-lam}
\bar z(\lambda) = \int_{0}^{z(\lambda)} \frac{dz'}{\sqrt {f(z')}}
\end{equation}

From this equation it is clear that $\bar z \ge 0$. Now for any finite value of $z$ the integral is convergent and $\bar z < \infty$. As $\lambda\rightarrow\infty$, $\bar z(\lambda)\rightarrow\infty$ which follows from (\ref{intcurve}). So $z(\lambda)\rightarrow\infty$ as $\lambda\rightarrow\infty$ otherwise $\bar z$ will attain infinite value for finite $z$ which is not possible. It is also true that as $\lambda\rightarrow -\infty$, $\bar z(\lambda)\rightarrow 0$. It follows from this that $z(\lambda)\rightarrow 0$ as $\lambda\rightarrow -\infty$ because the integrand in (\ref{bar-lam}) is positive definite.

 Let us now evaluate the expression in the limit of large and small $\lambda$. The expression for the $c$-function is, 

\begin{equation}\label{cfunc}
c(\lambda) = \frac{L}{4G_N} \frac{\bar z}{z} = \frac{L}{4G_N} \int_{0}^{1} \frac{d\alpha}{\sqrt {f(\alpha z(\lambda))}}
\end{equation}

where we have used $(\ref{bar})$. 

Now as $\lambda\rightarrow -\infty$, $z(\lambda)\rightarrow 0$ and using the fact that $f(0)=1$, we recover,

\begin{equation}
c(-\infty) = \frac{L}{4G_N} = \frac{c_{UV}}{6}
\end{equation}

Similarly, as $\lambda\rightarrow\infty$, $z(\lambda)\rightarrow\infty$ and using the fact that $f(\infty) = \frac{L^2}{L_{IR}^2}$, we recover,\footnote{If the reader is not comfortable with the fact that in (\ref{cfunc}) the lower limit of integration is $0$, one can replace it with a small number $\epsilon$ satisfying $0<\epsilon << 1$ and take the limit $\epsilon\rightarrow 0$ at the end. The answer will be the same, because there is no divergence in the integral anywhere.}

\begin{equation}
c(\infty) = \frac{L_{IR}}{4G_{N}} = \frac{c_{IR}}{6}
\end{equation}

Since $c'(\lambda)\le 0$ we conclude that $c(\lambda)$ monotonically decreases from the UV central charge to the IR central charge and $c_{IR}<c_{UV}$.

For a gapped system $f(\infty)$ blows up and we get $c_{IR}=0$. This is caustic formation. So caustic formation in the interior corresponds to a gapped system in the IR. 

\subsection{Arbitrary Dimension}

For $AdS_{d+1}$ the same thing goes through, except that the induced metric on $\bar\Sigma$ becomes,

\begin{equation}
ds_{ind}^2 = L^{2}\Big(\frac{\bar z}{z}\Big)^{2} ds^{2}_{H^{d-1}}
\end{equation}

where $ds_{H^{d-1}}^2$ is the metric on the unit hyperbolic space $H^{d-1}$ given by,

\begin{equation}
ds^2_{H^{d-1}} = d\eta^2 + \sinh^2\eta \ d\Omega^2_{d-2}
\end{equation}
The coordinates on the hyperbolic space are comoving.

The volume element of the horizon slice is,

\begin{equation}
dV = L^{d-1} \Big(\frac{\bar z}{z}\Big)^{d-1} dV_{H^{d-1}}
\end{equation}

We can again write, 

\begin{equation}
\frac{d}{d\lambda} \Big(L^{d-1} \Big(\frac{\bar z}{z}\Big)^{d-1}\Big) \le 0
\end{equation}

Now define the density of the Bekenstein-Hawking entropy of the Causal horizon slice,

\begin{equation}
\frac{dS_{BH}}{dV_{H^{d-1}}} = c(\lambda)= \frac{L^{d-1}}{4G_N} \Big(\frac{\bar z}{z}\Big)^{d-1}
\end{equation}

This is our $c$-function. We can also write it as,

\begin{equation}
c(\lambda)= \frac{L^{d-1}}{4G_N} \Bigg(\int_{0}^{1}\frac{d\alpha}{\sqrt{f(\alpha z(\lambda))}}\Bigg)^{d-1}
\end{equation}

Using the same argument as in $d=2$ we get that $c(\lambda)$ monotonically decreases between the UV central charge and the IR central charge. Here also caustic formation corresponds to a gapped phase in the IR.

\subsection{Stationarity Of The $c$-Function}

In this section we will check the stationarity of the $c$-function. This reduces to the problem of showing that the following quantity,

\begin{equation}
\frac{d}{d\lambda}c(\lambda) \propto \frac{d}{d\lambda} \Big(\frac{\bar z}{z}\Big) = \frac{\bar z}{z} \int_0^{1} \Bigg(1 - \frac{\sqrt {f(z(\lambda))}}{\sqrt {f(\alpha z(\lambda))}}\Bigg) d\alpha
\end{equation}

goes to zero as $\lambda\rightarrow -\infty$ or $\lambda\rightarrow +\infty$. 

As $\lambda\rightarrow -\infty$, $z(\lambda)\rightarrow 0$ and $\frac{\bar z}{z} \rightarrow 1$. This shows that that the integral vanishes in the UV making the $c$-function stationary at the UV fixed point. 

As $\lambda\rightarrow +\infty$, $z(\lambda)\rightarrow +\infty$ and $\frac{\bar z}{z}\rightarrow \frac{L_{IR}}{L}$. This makes the integral zero and the $c$-function stationary at the IR-fixed point. Here we have made use of the fact that as $z\rightarrow \infty$, $f(z)\rightarrow \frac{L^2}{L_{IR}^2}$ and as $z\rightarrow 0$, $f(z)\rightarrow 1$.

\section{Entanglement slicing}
Let us go back to the case of $AdS_3$ and choose a different slicing. We parametrize $\bar\Sigma$ by, 

\begin{equation}
x = t \cos\theta, \\ \bar z = t \sin\theta ,\\ t=t 
\end{equation}

By the same argument as in the previous case we get that $\theta$ is a comoving coordinate. The induced metric on the horizon is given by, 

\begin{equation}
ds_{ind}^2 = L^2\frac{t^2}{z^2} d\theta^2 = L^2 \Big(\frac{\bar z}{z}\Big)^{2} \frac{d\theta^2}{\sin^2\theta}
\end{equation}

Since $\theta$ is a comoving coordinate we can repeat the previous arguments and get back the same $c$-function. But we would like point out one thing. Suppose we are in pure $AdS$. In that case $\bar z = z$ and we get,

\begin{equation}
ds_{ind}^2 = L^2 \frac{d\theta^2}{\sin^2\theta}
\end{equation}

Now we can identify the quantity defined by, 

\begin{equation}
\int_{\epsilon}^{\pi-\epsilon} \frac{ds_{ind}}{4G_N}
\end{equation}

as the entanglement entropy of the UV-CFT. Here $\epsilon$ is a small dimensionless number representing a UV-cutoff on the comoving coordinate $\theta$. This is not the more conventional expression because we have chosen to cutoff the theory in a way which is compatible with the fact that in pure $AdS$ the future causal horizon is a Killing horizon. The usual holographic cutoff procedure uses a hard cutoff on the $z$ coordinate. The cutoff that we have used when translated to the $z$ coordinate becomes a space-time dependent cutoff in field theory. But this is a perfectly valid cutoff from the bulk gravity perspective. The fact that the EE does not depend on the system size in this scheme is the statement that there is no RG-flow in a CFT. The $c$-function that we have written down interpolates monotonically between the UV entanglement entropy and the IR entanglement entropy although in general it has no interpretation as EE at the middle of the RG-flow.

Before we conclude this section we would like to mention that with this particular choice of slicing the total entropy associated with a space-like slice can be identified with the Causal Holographic Information as defined in \cite{Hubeny:2012wa}. \footnote{I would like to thank Mukund Rangamani for pointing this out to me.} So the density of causal holographic information in a domain-wall geometry is a holographic $c$-function. 

\subsection{Linearized Second Law And Entropic c-Theorem}

The linearized first law for AdS-Rindler horizons has been shown to follow from the first law of entanglement entropy in the boundary field theory \cite{Lashkari:2013koa, Faulkner:2013ica}. There is an interesting special case when one can interpret the entropic $c$-theorem as the linearized second law for causal horizons. This is the case when the two fixed points are very close to each other so that we can do linearized calculation. This was done in \cite{Liu:2012eea} and the authors found that to linear order the renormalized EE is a good $c$-function for any space-time dimension. We think that the reason for this is the following. 

To prove entropic c-theorem, roughly speaking, one compares the EE of spherical regions of varying sizes \cite{Casini:2004bw, Casini:2012ei}. For example say we are working with a $(2+1)$ dimensional field theory. We consider a family of disks of varying radius $R$ given by,

\begin{equation}
x_1^2 + x_2^2 = R^2
\end{equation}

Let us start with a CFT. Since we are considering time-independent situation, we can as well think of the family of disks of different $R$ as slicing of the future light-cone of the point $(0,0,0)$ by constant time hypersurfaces, i.e,

\begin{equation}\label{bndligh}
 -t^2 + x_1^2 + x_2^2 =0 , \\ t=R
\end{equation}

This slicing gives rise to a slicing of the bulk future light-cone of the boundary point $(0,0,0)$ by constant-time hypersurfaces. This can be seen as follows. The bulk future light-cone has equation,

\begin{equation}
-t^2 + x_1^2 + x_2^2 + z^2 =0
\end{equation}

If we slice it by the plane $t=R$ we get,

\begin{equation}
-t^2 + x_1^2 + x_2^2 + z^2 =0, \\ t=R
\end{equation}

This reduces to $(\ref{bndligh})$ when $z=0$. The leaves of the foliation of the bulk light-cone are nothing but the Ryu-Takayanagi minimal surfaces ending on disks and their area computes the EE of the UV-CFT \cite{Ryu:2006bv,Casini:2011kv, Lewkowycz:2013nqa}. Now we consider perturbation by a relevant operator, the only restriction being that IR fixed point is very close to the UV fixed point so that we can do linearized calculation. To linearized order the change in the area of the Ryu-Takayanagi minimal surface can be computed by evaluating the area of the unperturbed minimal surface in the perturbed metric. The shape change does not contribute to the linearized order because the unperturbed surface is a minimal surface. So to compute the EE of the perturbed theory for varying disk sizes, holographically we just need to compute the areas of the different slicings of the unperturbed future light-cone in the perturbed metric. To see if they are increasing or decreasing as a function of $R$ we need to look at the evolution of the area along the unperturbed future light-cone. Now the linearized second law does precisely this thing. It tells us that the area along the bulk future light-cone decreases monotonically. This is the linearized $c$-theorem. We have done this for three dimensional field theories, but it is clear that it is true for any dimension.  It is also clear that beyond linearized order the entropic $c$-theorem does not match with the causal horizon $c$-theorem. 

\section{discussion}

In this paper we have tried to argue that the dynamics of the causal horizons in $AdS$ caused by the breaking of the scaling isometry is dual to the RG flow in the field theory. As a non-trivial check we have shown that the $c$-theorem is dual to the second law of thermodynamics of these dynamically evolving horizons. The Bekenstein-Hawking entropy density of these causal horizons play the role of $c$-function. So the number of degrees of freedom of the field theory is counted by the horizon density of Bekenstein-Hawking entropy. Since the slices of the causal horizon are noncompact, density is the right thing to use in this case. 

\subsection{Observer Dependence Of The c-Function}

One question is does the $c$-function depend on the choice of the bulk observer with which we associate the causal horizon. In the example we have studied, the $c$-function will not depend on our choice of the observer because the dual field theory is translationally invariant. It will be interesting to see what happens if we break translational invariance. Presumably the $c$-function will be replaced by a local quantity which will again show monotonic behavior.

\subsection{Higher Derivative Corrections And The Second Law}

Another important thing to consider is higher derivative corrections in the bulk. In Einstein gravity second law is the statement about the area increase of the spacelike slices of the future horizon. This is no longer true if we have higher derivative corrections to the Einstein theory in the bulk. The natural thing to consider in this situation is the Wald entropy \cite{Wald:1993nt,Iyer:1994ys}. But it is not known if Wald entropy satisfies the second law of thermodynamics \cite{Jacobson:1993vj,Jacobson:1995uq}. So Wald entropy cannot be used to produce a causal horizon $c$-function. In order to produce a causal horizon $c$-function we need an entropy functional which satisfies the second law. Recently a modified Wald functional has been shown to satisfy the linearized second law of black hole thermodynamics \cite{Bhattacharjee:2015yaa,Wall:2015raa, Dong:2013qoa, Camps:2013zua, Fursaev:2013fta, Bhattacharyya:2013jma}. If the UV and the IR fixed points are very close so that we can apply the linearized second law of thermodynamics to the dynamical causal horizon, then this expression defines the $c$-function. Now what about a general higher derivative theory of gravity ? A general higher derivative theory of gravity will not define a unitary boundary theory, but as long as the gravity theory satisfies the second law we can construct a causal horizon $c$-function using that. But if the higher derivative theory defines a unitary boundary theory then there will be $c$-theorem. This seems to imply that any unitary higher derivative theory of gravity satisfies the second law of causal horizon thermodynamics. It is far beyond the scope of this paper to say anything conclusive about it. But it is satisfying that the existence of the classical second law for causal horizons is closely tied to the unitarity of the underlying theory.

\subsection{Quantum Correction}

It is tempting to propose that at least the leading $\frac{1}{N}$ correction of this picture will be captured by the generalized second law applied to causal horizons. It will be interesting to see how this works out. We would like to take note of the fact that the generalized second law has been proved for causal horizons in \cite{Wall:2011hj,Wall:2010cj}.

\section{Acknowledgments} 
I would like to thank Sumit Das, D. Freedman, Simeon Hellerman, Charles Melby-Thompson, Djordje Radicevic, Steve Shenker and Sandip Trivedi for discussion and Mukund Rangamani for correspondence on related matter. I would like to thank especially Daniel Freedman for pointing out to me the importance of the stationarity property of the $c$-function. I would also like to thank the string theory group at Stanford University (SITP) for their hospitality during the completion of this project. This work was supported by World Premier International Research Center Initiative (WPI), MEXT, Japan.

\end{document}